\documentclass{article}
\usepackage{frascatiphys}
\usepackage{graphicx}
\begin{document}
\title{ 
Diphoton elastic scattering in UPC at smaller $W_{\gamma\gamma}$
}
\author{
Antoni Szczurek       \\
{\em Institute of Nuclear Physics PAN, Krak\'ow and University of Rzesz\'ow} \\
%Mariola K{\l}usek-Gawenda        \\
%{\em Institute of Nuclear Physics PAN, Krak\'ow} \\
}
\maketitle
\baselineskip=11.6pt
\begin{abstract}
We present a study of photon-photon scattering for $W_{\gamma\gamma} <
5$ GeV.
We extend earlier calculations of this cross section for 
$W_{\gamma\gamma}>$ 5 GeV into the low mass range
where photoproduction of the pseudoscalar mesons 
$\eta$(548), $\eta^{'}(958)$ and other mesonic resonances contribute 
to the two-photon final states.
We consider the dominant background of the two photon final state  
which arises from $\gamma \gamma$ decays of photoproduced 
$\pi^{0} \pi^0$-pairs.
We discuss how to reduce the background by imposing cuts on
different kinematical variables.
We present results for ALICE and LHCb kinematics.
\end{abstract}
\baselineskip=14pt
%

%----------------------------
\section{Introduction}
%----------------------------

First evidence of diphoton measurements in ultra-peripheral heavy-ion
collisions has been reported by the ATLAS and CMS Collaborations 
\cite{Aaboud:2017bwk,Sirunyan:2018fhl}. These data are,
however, restricted to photon-photon invariant masses W$_{\gamma\gamma} >$
5 and 6 GeV for the CMS and ATLAS analyses, respectively.
ATLAS comparison of its experimental results to the predictions
from Ref.\cite{Klusek-Gawenda:2016euz} show a reasonable agreement.
Our result is also consistent with the CMS data
\cite{Sirunyan:2018fhl}.

In our recent paper \cite{low-energy} we examined the possibility  of
measuring photon-photon scattering in ultra-peripheral heavy-ion collisions
at the LHC for $W_{\gamma\gamma}< 5$~GeV.
At lower diphoton masses, photoproduction of meson resonances plays a 
significant role
in addition to the Standard Model box diagrams \cite{Lebiedowicz:2017cuq},
as well as double photon fluctuations into light vector mesons 
\cite{Klusek-Gawenda:2016euz} or two-gluon exchanges
\cite{Klusek-Gawenda:2016nuo} may be imortant.
 
In our recent study we considered also background 
from the $\gamma \gamma \to \pi^0 (\to \gamma \gamma) \pi^0 
(\to \gamma \gamma)$ process measured e.g. by the Belle \cite{Uehara:2009cka} 
and Crystal Ball \cite{Marsiske:1990hx} collaborations. 
In Ref.~\cite{Klusek-Gawenda:2013rtu} a multi-component model,
which describes the Belle and Crystal Ball $\gamma \gamma \to \pi^0 \pi^0$ 
data, was constructed.

%--------------------------------------------
\section{Sketch of the formalism}
%--------------------------------------------

\begin{figure}[!h]
	(a)\includegraphics[scale=0.25]{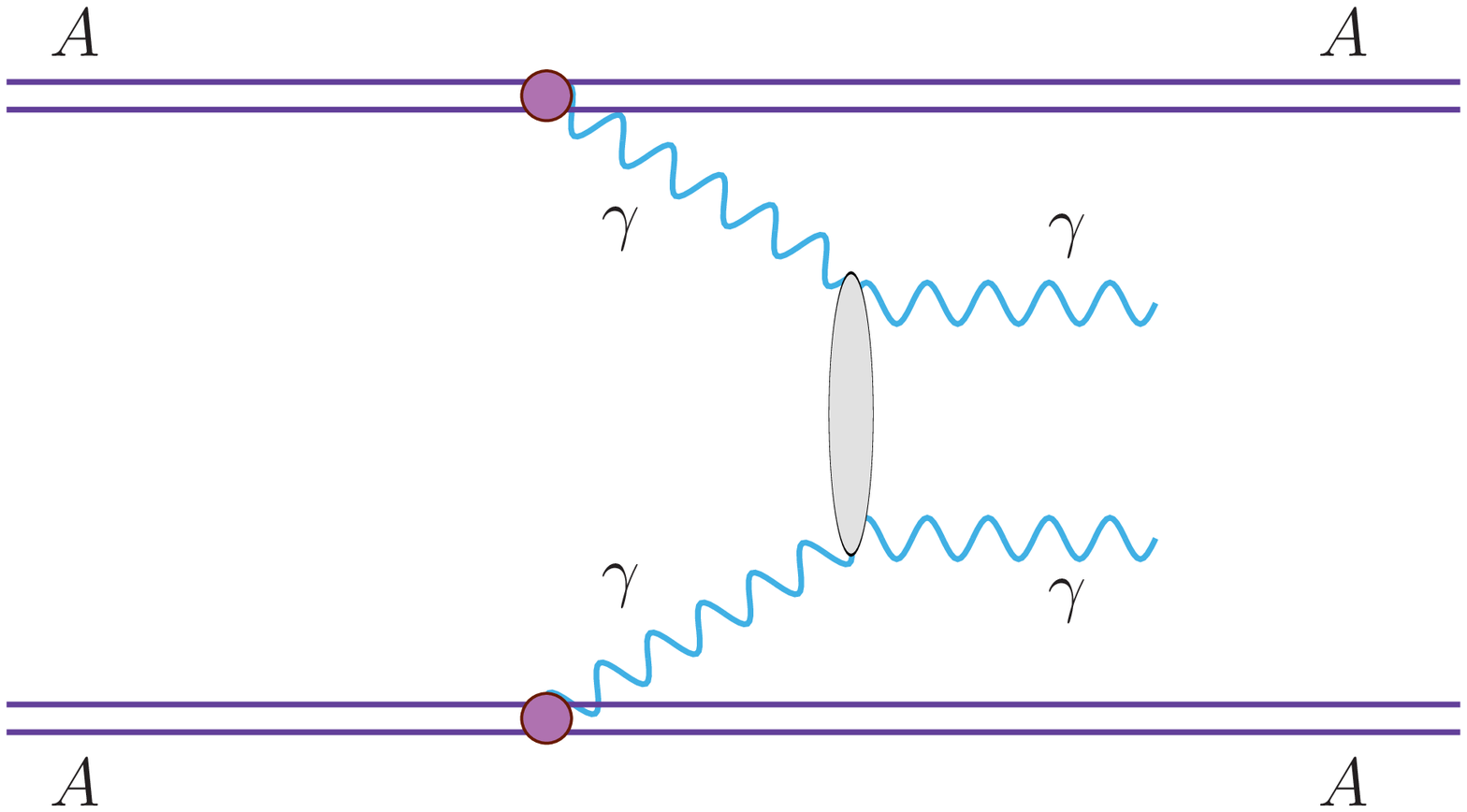} %\hspace*{1.cm}
	(b)\includegraphics[scale=0.25]{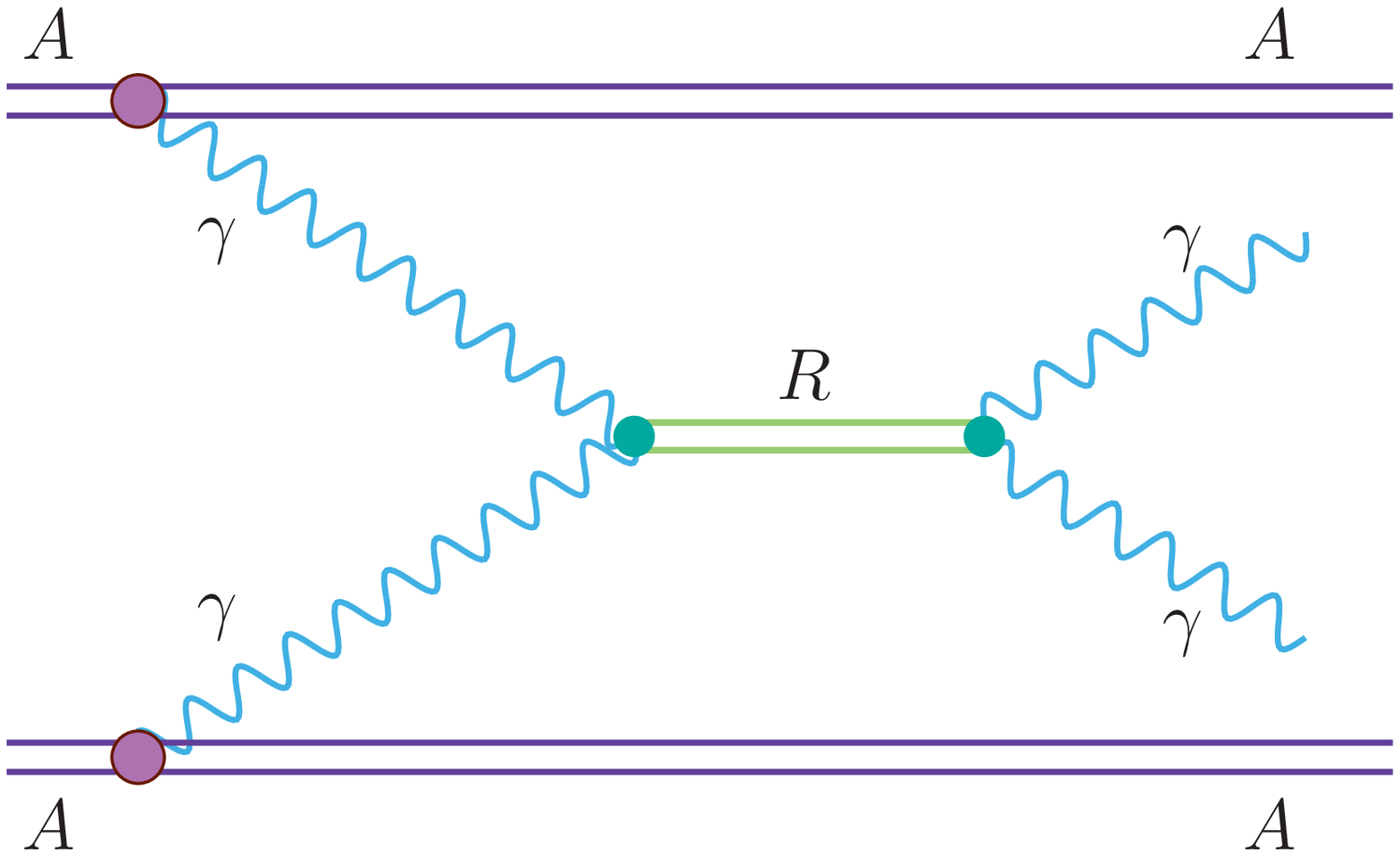}
	(c)\includegraphics[scale=0.25]{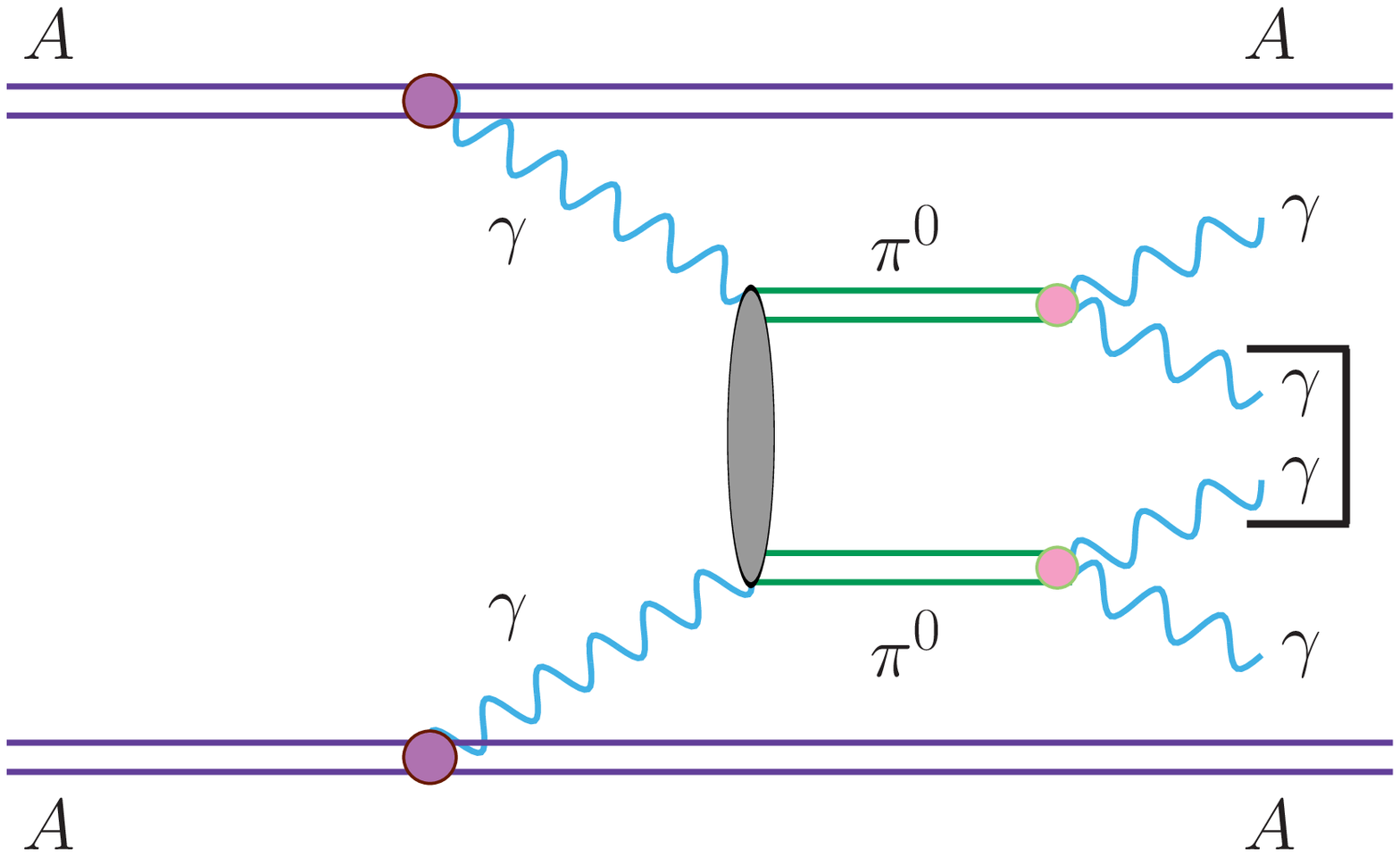}
	\caption{The continuum $\gamma \gamma \to \gamma \gamma$ scattering (a),
		$\gamma\gamma \to$ resonances $\to \gamma\gamma$ (b), 
		and the background mechanism (c).}
	\label{fig:diagrams}
\end{figure}

In Fig.~\ref{fig:diagrams} we illustrate the signal 
($\gamma \gamma \to \gamma \gamma$ scattering) which we take to be 
the dominant box mechanism (see \cite{Klusek-Gawenda:2016euz}).
Panel (b) shows a diagram for $s$-channel $\gamma\gamma \to $
pseudoscalar/scalar/tensor resonances which also 
contributes to the
$\gamma \gamma \to \gamma \gamma$ process.
We also show (diagram (c)) the $\gamma\gamma \to \pi^0\pi^0$ process, 
which leads to what we consider as the dominant background
 when only one photon from each $\pi^0 \to \gamma\gamma$
decay is detected. 

%---------------------------------------------------
%---------------------------------------------------

In our equivalent photon approximation (EPA) approach 
in impact parameter space,
the phase space integrated cross section for $A_1A_2 \to A_1 A_2 X_1
X_2$ reaction is expressed through, the five-fold integral
\begin{eqnarray}
\sigma_{A_1 A_2 \to A_1 A_2 X_1 X_2}\left(\sqrt{s_{A_1A_2}} \right) &=&
\int \sigma_{\gamma \gamma \to X_1 X_2} 
\left(W_{\gamma\gamma} \right)
N\left(\omega_1, {\bf b_1} \right)
N\left(\omega_2, {\bf b_2} \right)  \, S_{abs}^2\left({\bf b}\right) \nonumber  \\ 
& \times &
\mathrm{d}^2 b \, \mathrm{d}\overline{b}_x \, \mathrm{d}\overline{b}_y \, 
\frac{W_{\gamma\gamma}}{2}
\mathrm{d} W_{\gamma\gamma} \, \mathrm{d} Y_{X_1 X_2} \;,
\label{eq:EPA_sigma_final_5int}
\end{eqnarray} 
where $X_1 X_2$ is a pair of photons or neutral pions.
$W_{\gamma\gamma}=\sqrt{4\omega_1\omega_2}$
and $Y_{X_1 X_2}=\left( y_{X_1} + y_{X_2} \right)/2$ 
are invariant mass and rapidity of the outgoing $X_1 X_2$ system. 
The energy of the photons is expressed through 
$\omega_{1/2} = W_{\gamma\gamma}/2 \exp(\pm Y_{X_1 X_2})$.
${\bf b_1}$ and ${\bf b_2}$ are impact parameters 
of the photon-photon collision point with respect to parent
nuclei 1 and 2, respectively, 
and ${\bf b} = {\bf b_1} - {\bf b_2}$ is the standard impact parameter 
for the $A_1 A_2$ collision.
The absorption factor $S_{abs}^2\left({\bf b}\right)$ assures UPC 
implying that the nuclei do not undergo nuclear breakup.
The photon fluxes 
($N\left(\omega_i, {\bf b_i} \right)$) are 
expressed through a nuclear charge form factor of the nucleus. 
In our calculations we use a realistic form factor which is a 
Fourier transform of the charge distribution in the nucleus.
More details can be found e.g. in \cite{KlusekGawenda:2010kx}.

%-------------------------
\section{Results}
%-------------------------

In Table 1 we show the total cross sections in nb for different 
contributions to the diphoton final state.
The cross sections are given in two ranges of di-photon invariant masses
both for ALICE and LHCb acceptances. The rapidity coverage of ALICE
is $|\eta_{\gamma}| <$ 0.9 and LHCb 2 $< \eta_{\gamma} <$ 4.5.

%---------------------------------------------------------------------------
		\begin{table}[!h]
\centering
\caption{Total nuclear cross section in nb for the Pb+Pb collisions for 
$\sqrt{s}_{NN}$ = 5.02 TeV.}
			\begin{tabular}{|l|r|r|r|r|}
				\hline
				Energy   &  \multicolumn{2}{c|}{$W_{\gamma\gamma} = (0-2)$ GeV}  
				&  \multicolumn{2}{c|}{$W_{\gamma\gamma}>$ 2 GeV}   \\ \hline
				Region	& ALICE		& LHCb		& ALICE		& LHCb	\\ \hline \hline
				boxes       &   4 890 	& 3 818		& 146		& 79  	\\ \hline
				$\pi^0\pi^0$ bkg  & 135 300	& 40 866	& 46 		& 24	\\ \hline
				$\eta$		& 722 573	& 568 499	& 			&		\\ \hline
				$\eta'(958)$&  54 241	&  40 482	& 			&		\\ \hline
				$\eta_c(1S)$&			&			& 9			& 5		\\ \hline
				$\chi_{c0}(1P)$& 		&			& 4			& 2		\\ \hline
				$\eta_c(2S)$&			&			& 2			& 1		\\ \hline	 
			\end{tabular}
		\end{table}
%-------------------------------------------------------------------------

\begin{figure}[!h]
\centering
\vspace*{-.4cm}
\includegraphics[scale=0.235]{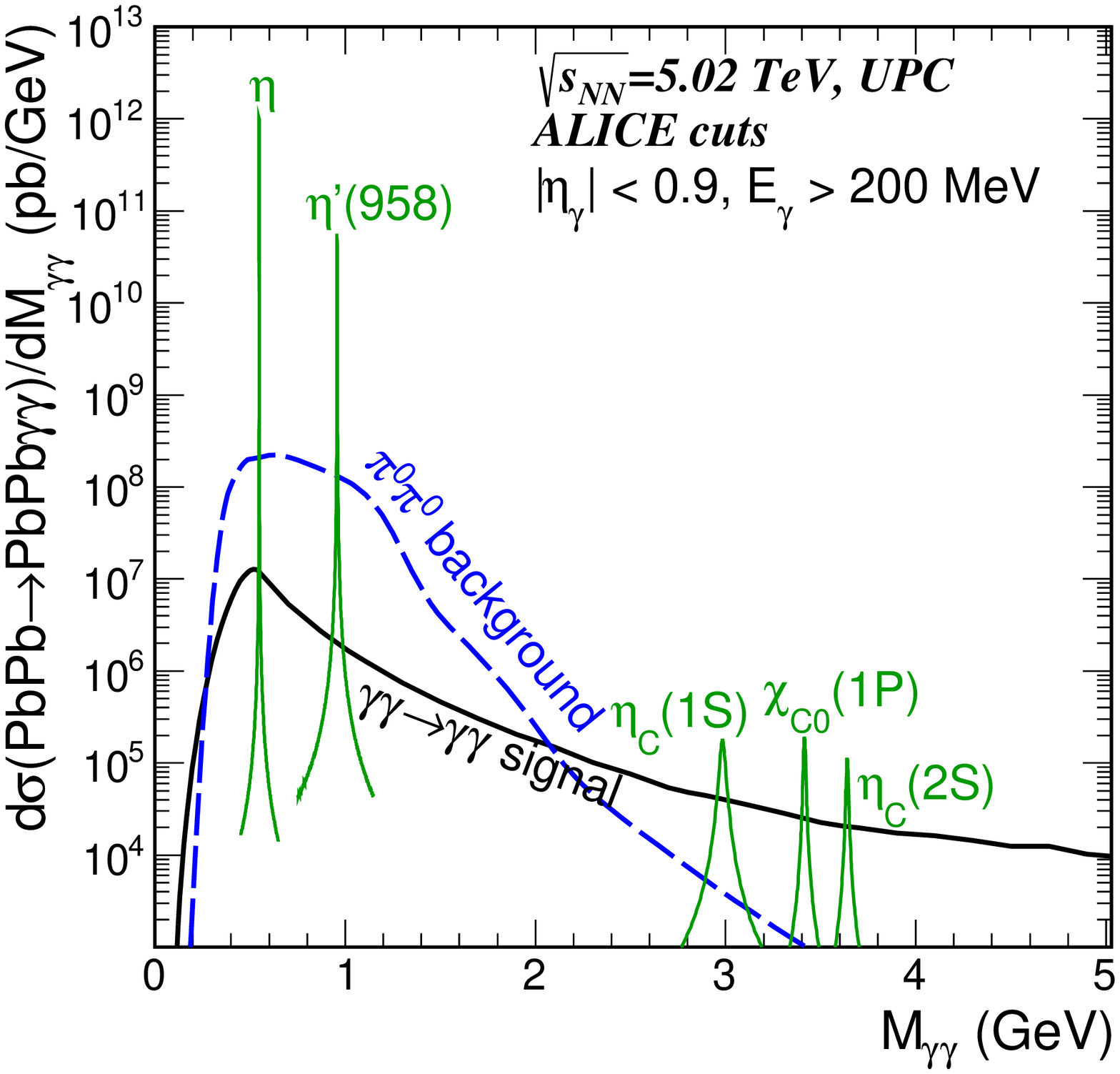}
\includegraphics[scale=0.235]{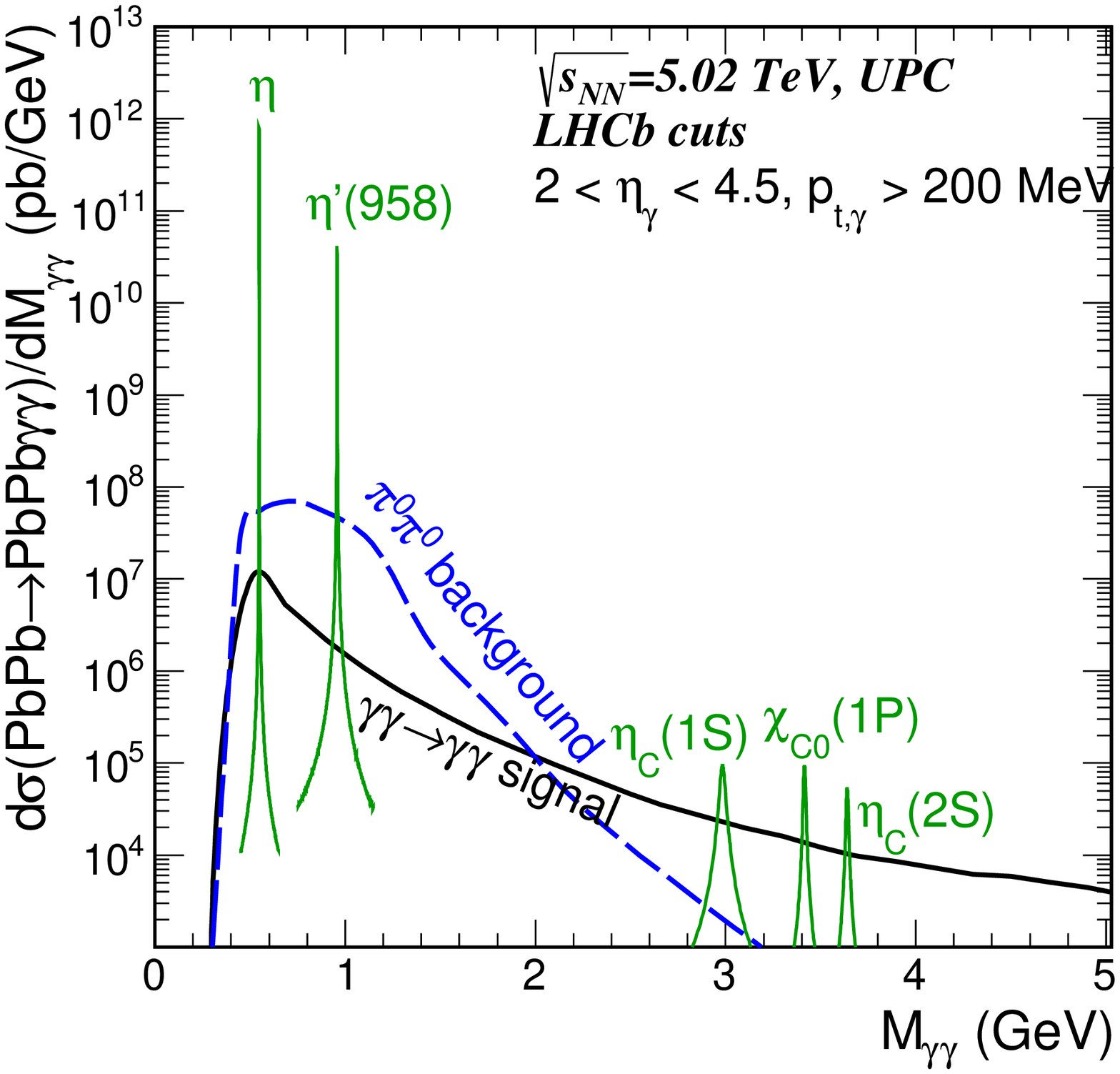}
\caption{Diphoton invariant mass distribution for ALICE (left panel) and
  LHCb (right panel) kinematical conditions.
}
\label{fig:dsig_dM}
\end{figure}

In Fig.\ref{fig:dsig_dM} we show distribution in the diphoton invariant
mass, separately for the ALICE (left panel) and LHCb (right panel)
kinematics. As a signal (solid line) here we included only fermionic box
contributions. One can observe sharp peaks corresponding to the
s-channel exchanges of many resonances specified in the figure.
In the calculation presented in this figure only cuts on photon
rapidities and transverse momenta were imposed. No experimental
resolutions were included, therefore the peaks corresponding
to mesons are fairly sharp. The dashed line represents the background
due to incomplete registration of the $\pi^0 \pi^0$ channel.
The background is rather small above $M \sim$ 2.5 GeV. Therefore a
measurement of $\gamma \gamma \to \gamma \gamma$ scattering for 
subchannel energies $W >$ 3 GeV should be possible. We remind that ATLAS
and CMS could measure $\gamma \gamma \to \gamma \gamma$ scattering only 
for energies larger than 6 and 5 GeV, respectively.

%-----------------------------------------------------------------------------------------------
%-----------------------------------------------------------------------------------------------

\begin{figure}[!h]
\centering
\vspace*{-.4cm}
\includegraphics[scale=0.235]{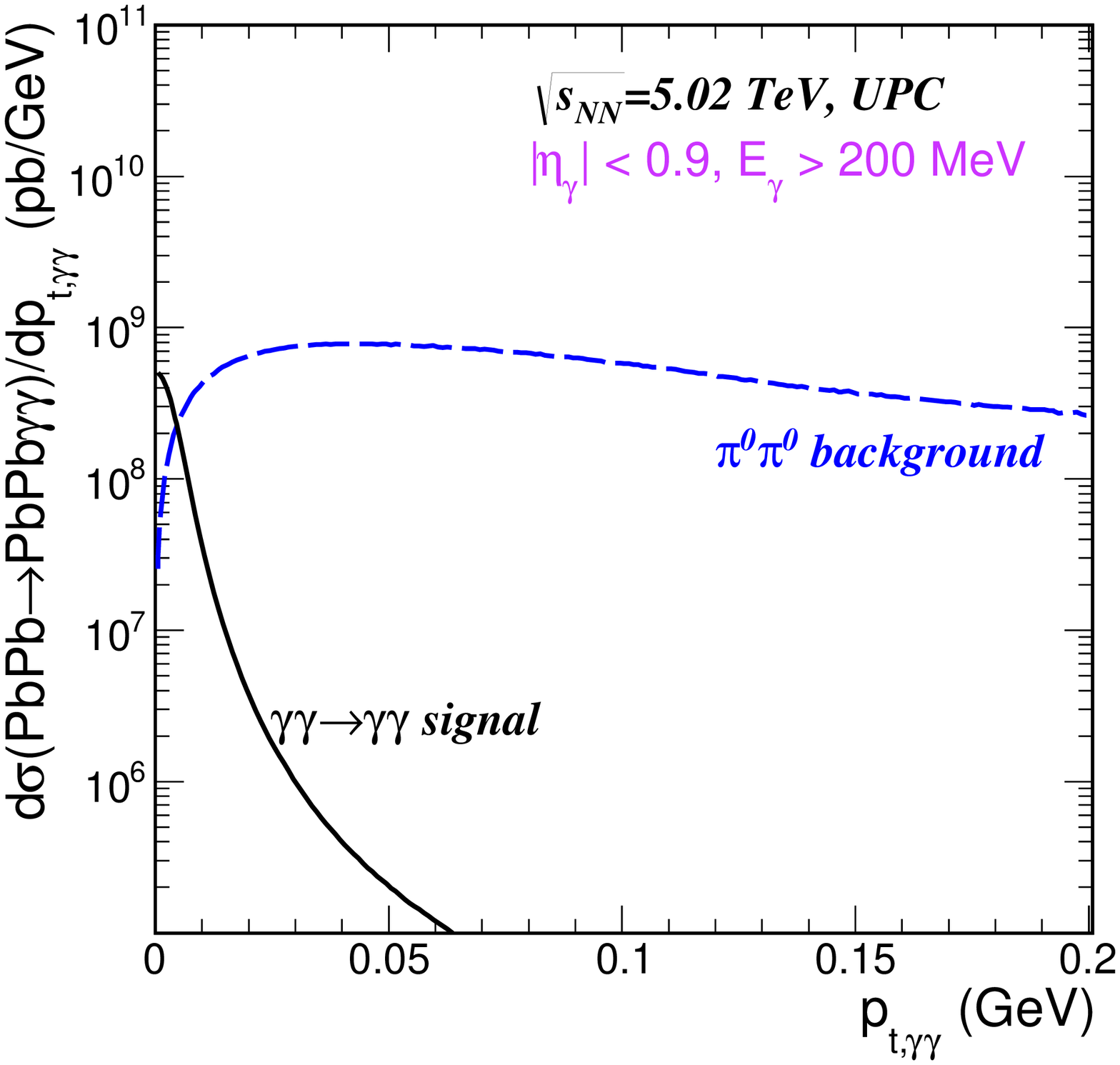}
\includegraphics[scale=0.235]{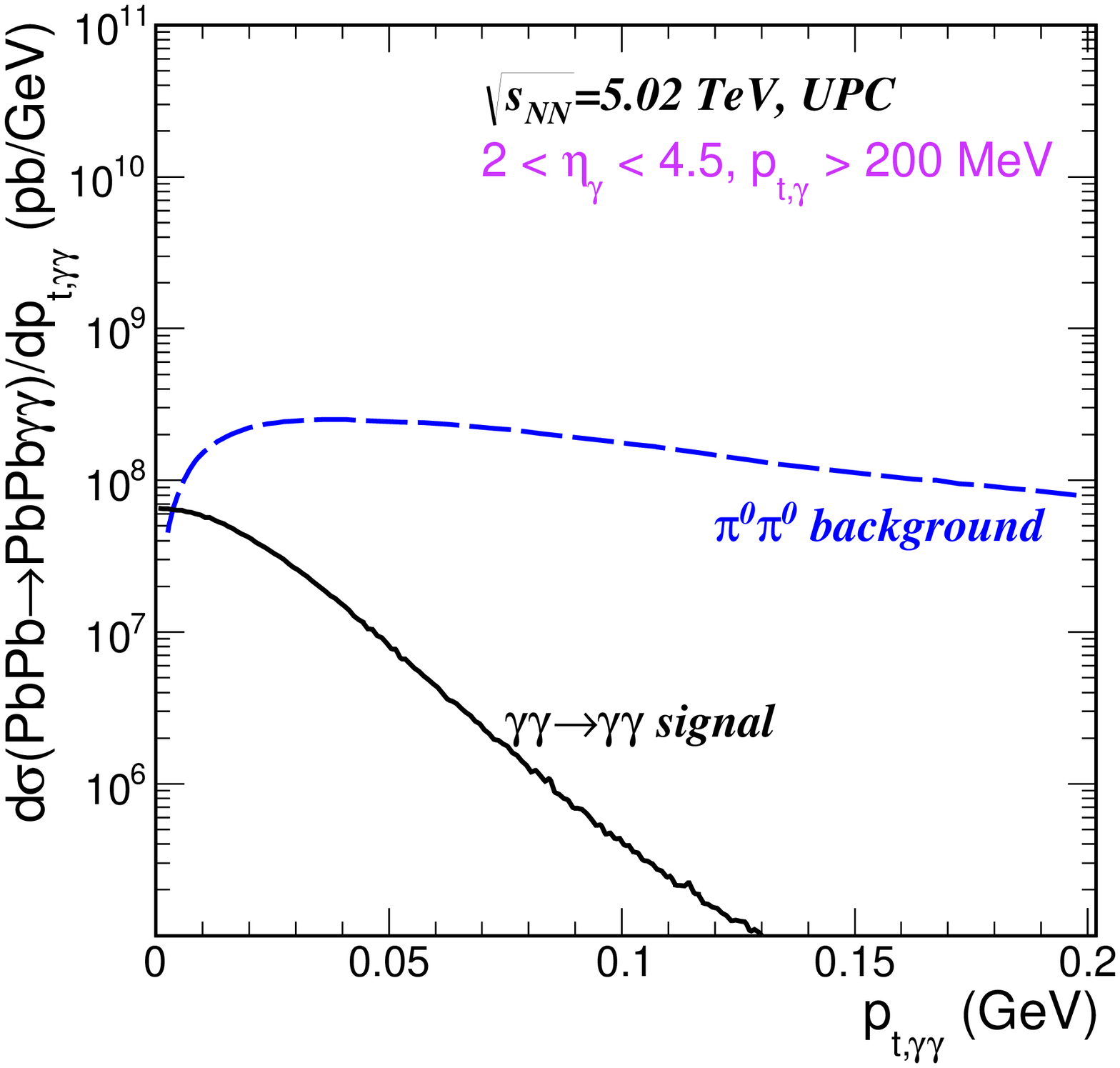}
\caption{Transverse momentum distribution of the diphoton pair
for ALICE (left) and LHCB (right) kinematics.}
\label{fig:dsig_dptsum}
\end{figure}

How to reduce the unwanted background ?
As an example in Fig.\ref{fig:dsig_dptsum} we show distribution in
transverse momentum of the diphoton pair. The solid line represents 
the signal and the dashed line the background evaluated separately for 
the ALICE and LHCb experimental conditions. The smearing in 
$p_{t,\gamma \gamma}$ is caused by finite experimental energy resolution
included in this calculation.
It is clear that imposing 
extra cuts on transverse momenta of the pair of photons one can get 
rid off the unwanted background. Several other possibilities
how to reduce the background were considered in our original 
paper \cite{low-energy}.

%---------------------------------------------------------------
%----------------------------------------------------------------

\begin{figure}[!h]
\centering
\vspace*{-.cm}
\includegraphics[scale=0.3]{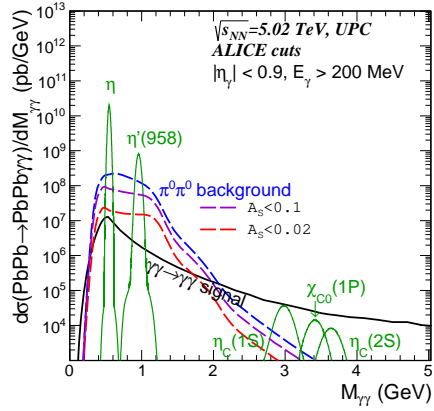}
\caption{Diphoton invariant mass for the ALICE conditions including
experimental energy resolution. Here we show also effect of cuts on
scalar asymmetry of outgoing photon transverse momenta.
}
\label{fig:dsig_dM_resolution}
\end{figure}

The effect of energy resolution on diphoton invariant mass spectra 
is shown in Fig.\ref{fig:dsig_dM_resolution}
for the ALICE case. Here we show also the effect of imposing a cut
on so-called scalar asymmetry of outgoing photons (see \cite{low-energy}).

%----------------------------------------------------------------
%-----------------------------------------------------------------

\begin{figure}[!h]
\centering
\vspace*{-.cm}
\includegraphics[scale=0.235]{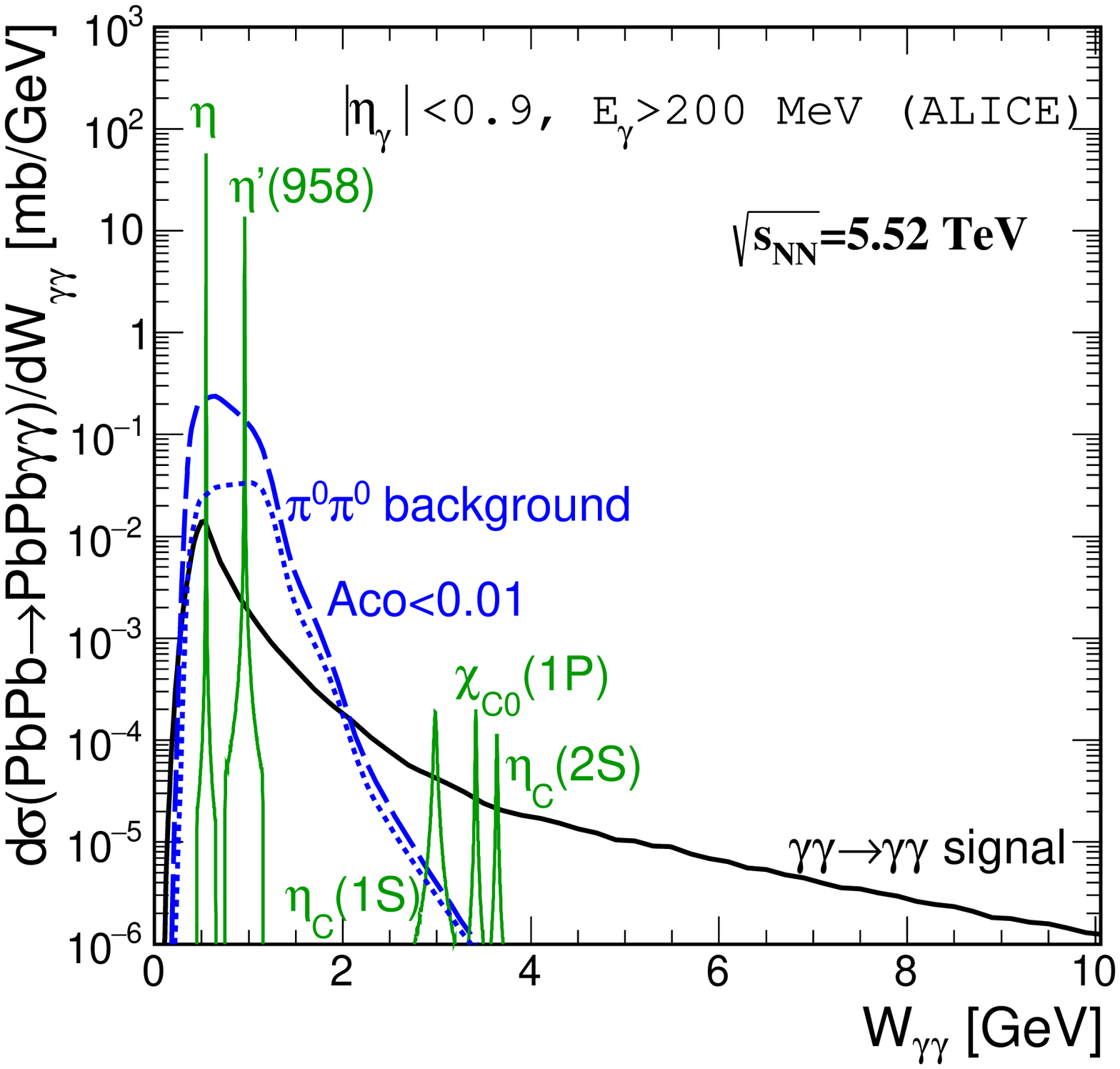}
\includegraphics[scale=0.235]{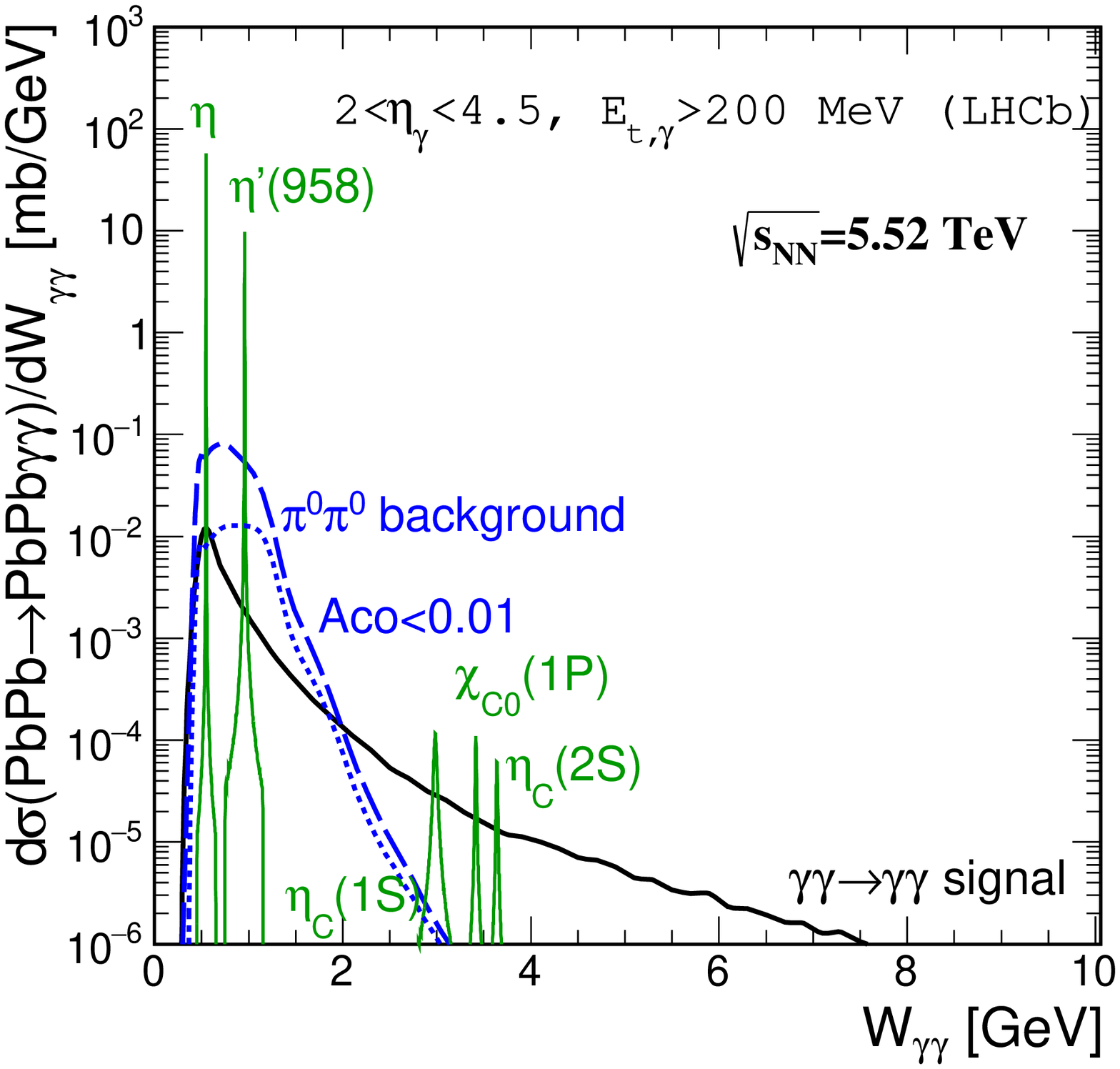}
\caption{Diphoton invariant mass distribution for ALICE (left) and LHCb (right)
  kinematics. Here a cut on acoplanarity is imposed.}
\label{fig:dsig_dM_ALICE+LHCb_acoplanarity}
\end{figure}

Acoplanarity is another variable which can be used to reduce the
$\pi^0 \pi^0$ background. In Fig.\ref{fig:dsig_dM_ALICE+LHCb_acoplanarity}
we demonstrate the effect of limiting acoplanarity range. Even with 
drastic cuts on acoplanarity it is very difficult to reduce 
the $\pi^0 \pi^0$ background.

%-----------------------------------------------------------------
%------------------------------------------------------------------

\begin{figure}[!h]
\centering
\vspace*{-.cm}
\includegraphics[scale=0.235]{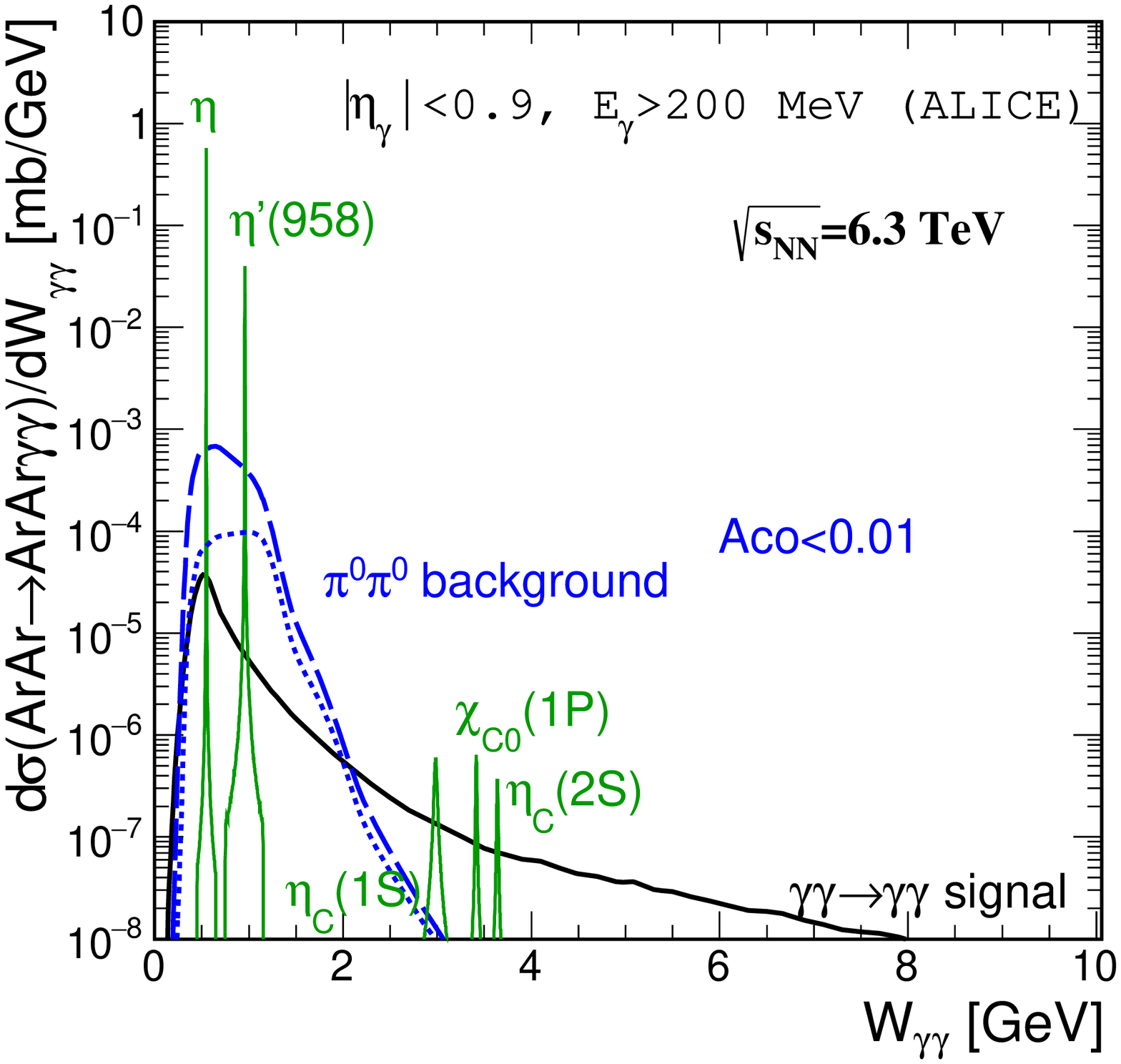}
\includegraphics[scale=0.235]{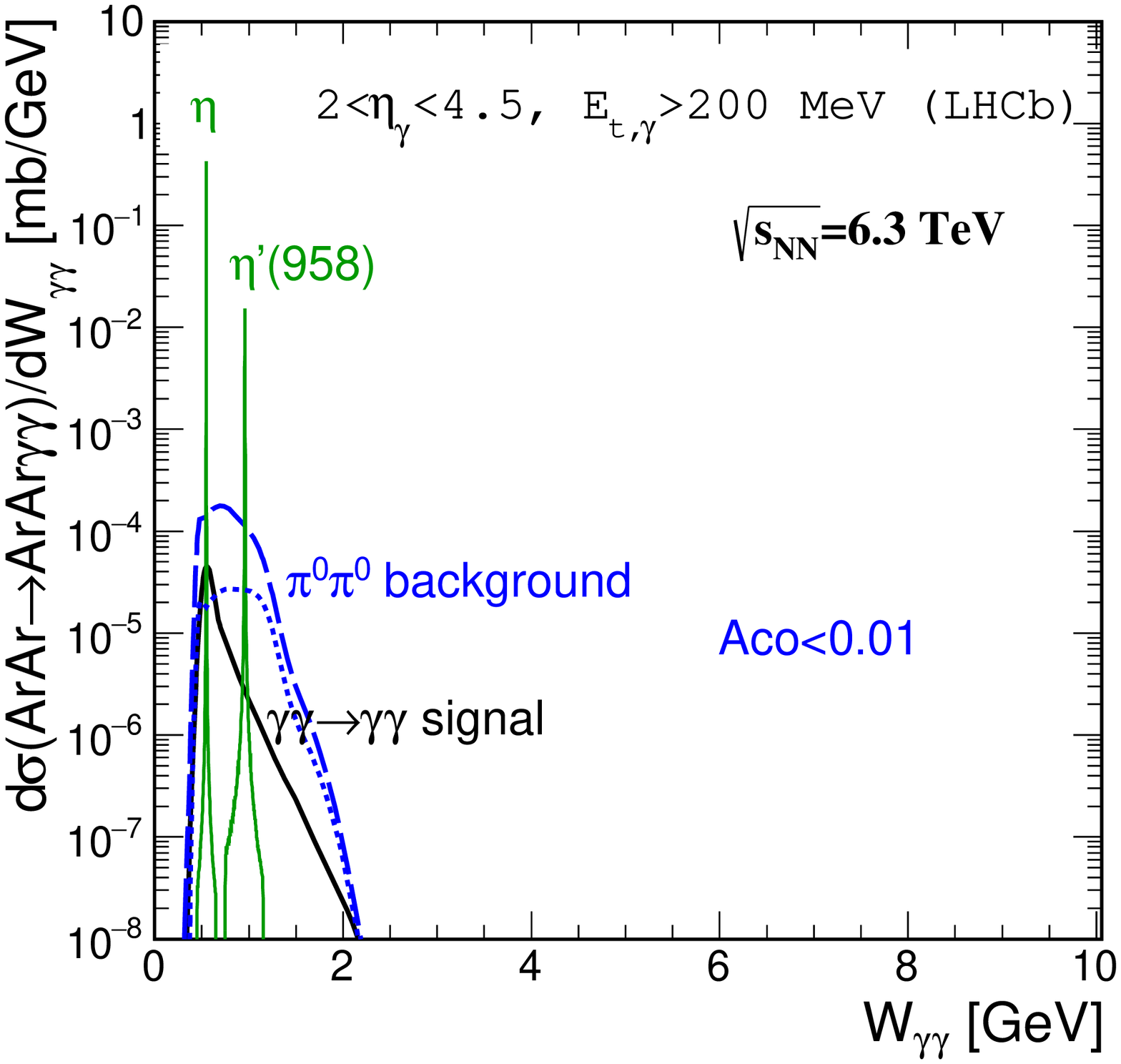}
\caption{Diphoton invariant mass distributions for Ar + Ar collisions 
with the acoplanarity cut imposed.}
\label{fig:dsig_dM_ALICE+LHCb_acoplanarity_ArAr}
\end{figure}

In Fig.\ref{fig:dsig_dM_ALICE+LHCb_acoplanarity_ArAr} we show a similar
result for Ar + Ar collisions. The situation looks similar as for 
Pb + Pb collisions. Although the cross section for Ar + Ar collisions 
is much smaller than that for Pb + Pb collisions the reaction can be
very useful due to higher integrated luminosity and in the consequence
higher counting rate.

%-----------------------------------------------------------------
%------------------------------------------------------------------

%--------------------------
\section{Conclusions}
%--------------------------

Here we have considered the possibility to study elastic $\gamma \gamma \to$
$\gamma \gamma$ scattering in the diphoton mass range
$W_{\gamma\gamma} <$ 5 GeV at the LHC using ALICE or LHCb detectors.
Our results show that the contributions of the pseudoscalar resonances
$\eta$(548), $\eta^{'}(958)$ are clearly visible on top of the diphoton
mass continuum arising from fermion loop diagrams.
We have made first predictions for cross sections 
as a function of diphoton mass for typical acceptances in rapidity 
and transverse momentum of the ALICE and LHCb experiments.
The evaluation of counting rates needs, however, Monte Carlo simulations
which take into account detailed acceptances and realistic responses of 
the detectors used for measuring the two-photon final states.

In addition to the signal ($Pb Pb \to Pb Pb \gamma \gamma$)
we considered also the background dominated by the 
$Pb Pb \to Pb Pb \pi^0 \pi^0$
reaction when only two out of the four decay photons in the final state 
are registered. This background can be reduced by imposing cuts on scalar
and vector asymmetry of transverse momentum of the two photons,
acoplanarity, etc.
In Ref.\cite{small-energy} we showed also that cuts on the sum of photon
rapidities (or the rapidity of the diphoton system) can additionally 
be used to reduce the background.

%---------------------------------
\section{Acknowledgements}
%---------------------------------

I am indebted to Mariola K{\l}usek-Gawenda, Ronan McNuclty and Rainer
Schicker for collaboration on the topic presented here.
This work was supported by the Polish National Science Center 
grant DEC-2014/15/B/ST2/02528. 

%------------------------
%\section{References}
%------------------------
%

\end{document}